\documentclass[conference]{IEEEtran}
\IEEEoverridecommandlockouts
\usepackage{cite}
\usepackage{amsmath,amssymb,amsfonts}
\usepackage{algorithmic}
\usepackage{graphicx}
\usepackage{textcomp}
\usepackage{xcolor}
\usepackage{url}
\usepackage{hyperref}
\usepackage{orcidlink}
\def\BibTeX{{\rm B\kern-.05em{\sc i\kern-.025em b}\kern-.08em
    T\kern-.1667em\lower.7ex\hbox{E}\kern-.125emX}}
\begin{document}

\title{Direct phase encoding in QAOA: Describing combinatorial optimization problems through binary decision variables\\
\thanks{This work has been supported by the Ministry of Economic Affairs Baden-Württemberg in the project SEQUOIA under project number 036-840012.}
}

\author{\IEEEauthorblockN{Simon Garhofer* \orcidlink{0000-0002-6857-3214}}
\IEEEauthorblockA{\textit{Department of Computer Science} \\
\textit{University of Tübingen}\\
Tübingen, Germany \\
simon.garhofer@uni-tuebingen.de}
\and
\IEEEauthorblockN{Oliver Bringmann}
\IEEEauthorblockA{\textit{Department of Computer Science} \\
\textit{University of Tübingen}\\
Tübingen, Germany \\
oliver.bringmann@uni-tuebingen.de}
}

\maketitle

\begin{abstract}
The Quantum Approximate Optimization Algorithm (QAOA) and its derived variants are widely in use for approximating combinatorial optimization problem instances on gate-based Noisy Intermediate Scale Quantum (NISQ) computers. Commonly, circuits required for QAOA are constructed by first reformulating a given problem as a Quadratic Unconstrained Binary Optimization (QUBO) problem. It is then straightforward to synthesize a QAOA circuit from QUBO equations.

In this work, we illustrate a more qubit-efficient circuit construction for combinatorial optimization problems by the example of the Traveling Salesperson Problem (TSP). Conventionally, the qubit encoding in QAOA for the TSP describes a tour using a sequence of nodes, where each node is written as a 1-hot binary vector. We propose to encode TSP tours by selecting edges included in the tour. Removing certain redundancies, the number of required qubits can be reduced by a linear factor compared to the aforementioned conventional encoding. 

We examined implementations of both QAOA encoding variants in terms of their approximation quality and runtime. Our experiments show that for small instances results are just as accurate using our proposed encoding, whereas the number of required classical optimizer iterations increases only slightly.
\end{abstract}


\section{Introduction}

Combinatorial optimization problems in general and the TSP in particular are \emph{NP-hard} problems, hence there are no algorithms known that return optimal solutions for this class of problems efficiently, i.e. in a feasible amount of time for growing problem instance sizes. In the worst case, algorithms that compute optimal solutions have to check every possible solution given in the problem statement, resulting in exponential runtime. In most practically relevant instances, approximation algorithms are used which compute solutions in considerably more favorable runtime at the cost of optimality \cite{tsp}.

Quantum computers are capable of finding optimal solutions to combinatorial optimization problems more quickly, however, the asymptotic runtime of all known algorithms is still exponential \cite{grover}\cite{minfind}. Moreover, in the current era of NISQ computers algorithms of such complexity will result in extremely noisy computations, making the output effectively unusable \cite{bittertruth}.

Hence, as long as superpolynomial run-time improvements for exact quantum algorithms remain unproven, approximation algorithms are considered a viable alternative. In this work, we concern ourselves with QAOA, which for the TSP in particular offers no asymptotic runtime advantages over classical approximation algorithms \cite{tsp}\cite{qaoap}. While there are precious few proven theoretical bounds for QAOA approximation ratios, especially for implementations using many layers, there is empirical evidence for better approximation ratios in practical applications. Further results would need to be taken from physical backends -- ideally providing many more qubits than as of today \cite{qaoa}\cite{qaoareview}\cite{boundsmaxcut}\cite{boundsrounds}.

An essential aspect of any QAOA circuit is how internal states represent candidate solutions. We refer to this as the (problem) encoding. A straightforward way of constructing such an encoding is to restate a given problem as a QUBO. This QUBO is then transformed into an Ising Hamiltonian, the ground state of which is determined using QAOA \cite{qubo}\cite{qaoa}.

Dependent on the specific problem, such encodings may end up wasteful in the number of qubits. In the TSP QAOA for instance, each city is represented by a 1-hot binary string. The order of these strings depicts the sequence of cities. Our goal is to reduce the number of qubits required by leveraging phase summation effects of the tensor product for tour length computation while maintaining the expressiveness of the encoding.

\subsection{Related work}

Since the TSP is a fundamental problem in classical computer science as well as ubiquitous in practical applications, it is widely covered in quantum computing research, too. State encodings in QAOA have been the subject of several works. In \cite{encodingbinary} the authors suggest replacing the 1-hot encoding with a binary encoding. Naturally, this reduces the number of required qubits from quadratic to linearithmic but leaves open a concrete circuit construction.

The authors of \cite{encodingoptimized} argue that simply changing the encoding to a binary representation disproportionately amplifies circuit depth, making it NISQ-unfriendly. They therefore present a hybrid encoding where the fraction of binary and 1-hot encoding can be chosen.

In \cite{encodingtradeoff} it is suggested to divide the QUBO subspace using two qubit arrays: $n_r$ register qubits are used to address $2^{n_r}$ different subsystems with $n_a$ qubits. Those $n_a$ qubits are then able to express arbitrarily complex correlations between classical variables.

\subsection{Contribution}

We describe a more qubit-efficient QAOA circuit construction for the TSP in Section \ref{methods}, focusing on the problem encoding. A mathematical proof of its correctness is given. Our encoding reduces the number of required qubits by a linear factor and removes cyclic redundancies.

Experimental results are shown in section \ref{experiments}. We find that our proposed TSP encoding returns good solutions in small problem instances at the same rate the conventional encoding does.

All required preliminaries are briefly covered in section \ref{prelims} with respective citations providing more detailed information.

\section{Preliminaries} \label{prelims}

Let $G = (V, E)$ be a complete directed graph with vertices $V, \vert V\vert=n$ and edges $E, \vert E\vert=n^2-n$ and let $c: E\rightarrow \mathbb{R}$ be a function that maps each edge to some cost. The TSP asks to find a tour that visits all vertices exactly once and returns to the origin vertex while the cost of the tour, determined by $c$, is minimal. \cite{tsp}. \\

QAOA is a hybrid algorithm for optimizing combinatorial optimization problems. The basic QAOA setup as described in \cite{qaoa} consists of a parametrized quantum circuit that is iteratively optimized using classical algorithms. In the quantum circuit an operator $C$ that implements the cost function is, alternately with a mixer operator $B$, applied to a uniform superposition of all basis states, dependent on angles $\gamma_1, ..., \gamma_p, \beta_1, ..., \beta_p$:
\begin{align*}
    \vert\gamma,\beta\rangle = e^{-i\beta_pB}e^{-i\gamma_pC}\hdots e^{-i\beta_1B}e^{-i\gamma_1C}\vert +\rangle
\end{align*}
The expected value $\langle\gamma,\beta\vert C \vert\gamma,\beta\rangle$ is then minimized/maximized by some classical optimization algorithm.

The procedure described in \cite{qaoap} extends QAOA by constructing the mixer in such a way that given problem boundary conditions are adhered to. Constraints, such as ``every city has to be visited exactly once'', are therefore guaranteed to be met, whereas in standard QAOA such constraints would have to be implemented via penalty terms in $C$.

The mixer construction is simplified in \cite{gqaoa}, resulting in so-called \emph{Grover mixers}. This modification necessitates a more complex initial state preparation, where instead of trivially preparing a single valid state (with respect to the boundary conditions), all such states are prepared in a uniform superposition. Moving circuit complexity from the mixer to the state preparation serves as a trade-off, effectively reducing circuit depth for large $p$ given that the initial state can be prepared efficiently. \\

Boundary conditions are enforced by only considering a proper subset of all $2^n$ states a register of $n$ qubits can be in as a valid encoding of some TSP tour. Mixers in the Alternating Operator Ansatz and thus Grover mixers transform any such state into another valid state, to some given degree $\beta_i$. Given that initially no invalid state was prepared, the circuit will satisfy all boundary conditions encoded in the mixers \cite{qaoap}\cite{gqaoa}.

Going further, we concern ourselves with Grover mixers (\emph{GQAOA}) for the TSP exclusively. Note though that our proposed encoding does not enforce a specific mixer type. \\

Conventionally, TSP tours in \cite{qaoa}, \cite{qaoap} and \cite{gqaoa} are encoded by describing a sequence of vertices. We name those vertices $v_i$ such that $V=\{v_0, ..., v_{n-1}\}$ and similarly an edge between $v_i$ and $v_j$ $e_{i,j}$. The cost of an edge is denoted as $c(e_{i,j})$.

A sequence of vertices $(\iota_1, \iota_2, ... \iota_n)$ describes a tour, where $\iota_1$ is visited first and is returned to after visiting $\iota_{n}$. In the conventional encoding, for every combination $(\iota_i,v_j),\ 1\leq i\leq n,\ 0\leq j\leq n-1$ there is a variable $z_{i,j}$. When $\iota_i = v_j$ (i.e. $v_j$ is visited in time step $i$), $z_{i,j} = 1$, otherwise $z_{i,j} = 0$. The TSP problem statement requires the following conditions to hold:

\begin{align*}
    \sum_j z_{i,j} = 1 & \ & \forall\ i = 1, ..., n \\
    \sum_i z_{i,j} = 1 & \ & \forall\ j = 0, ..., n-1
\end{align*}

Intuitively this means that in a single time step $i$ exactly $1$ vertex needs to be visited. Each vertex $v_j$ needs to be visited exactly once. \\

\begin{figure}[htbp]
\centerline{\includegraphics[width=0.35\textwidth]{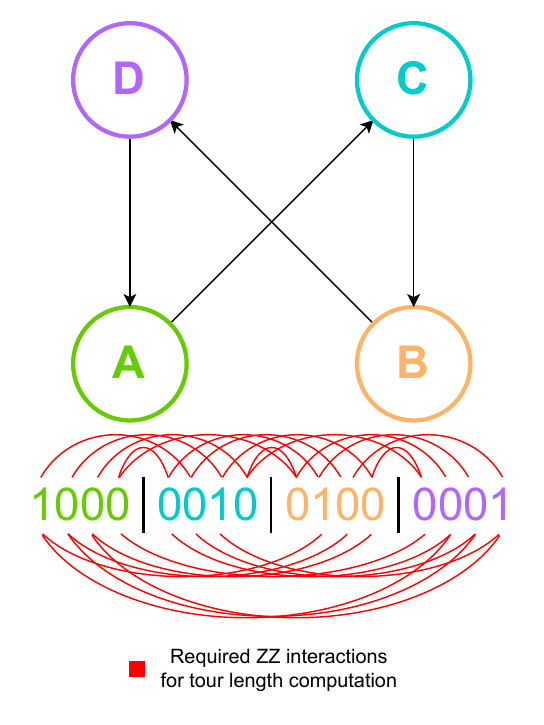}}
\caption{Example for the conventional encoding for the TSP in a QAOA setting. Vertices are 1-hot encoded and then written in the desired order. In order to determine the length of the encoded tour the involved edges need to be identified. This is done by checking the order in which all pairs of vertices appear in the tour using ZZ rotations.}
\label{EncodingOldSolo}
\end{figure}

To retrieve the tour length from this encoding in the phase separator all taken edges are computed and summed up using pair-wise ZZ rotations. This setup results in the function
\begin{align}
    C(\iota) = 4\cdot f(\iota) - (n-2) \cdot \sum_{j=1}^n \sum_{k=1}^n c(e_{j,k})
\end{align}
where $f(\iota)$ is the tour length of the ordering $\iota$ \cite{qaoap}. Figure \ref{EncodingOldSolo} depicts an example tour encoding for a tour through four nodes.\\

Each ZZ rotation requires two CNOT gates to implement, making the numerous use of RZZ gates unsuited for NISQ. Using the described encoding requires $\mathcal{O}(n^3)$ RZZ gates for each phase separator since the number of qubits scales quadratically with the number of vertices \cite{qaoap}.

Additionally, the conventional encoding fails to express a given tour uniquely since any sequence of vertices can be shifted cyclically $n$ times. In practice this means that $n$ different qubit states encoding a possibly optimal solution will compete for high measurement probabilities, essentially sabotaging each other. Especially in the presence of noise this makes it more likely for good solutions to be drowned out by noisy measurements. This can be remedied though by fixing one vertex as a starting point for all tours and making minor changes to the tour length function \cite{qaoap}.

\section{New encoding} \label{methods}

We first outline our proposed TSP encoding on a high level before giving a detailed mathematical description. \\

The main idea of our proposed encoding is to not encode the order of cities but rather the selection of edges in the input graph itself. Each edge is represented by a single qubit. If the qubit is in state $\vert0\rangle$, the corresponding edge is not part of the tour. Otherwise, it is.

Naturally, this approach simplifies the computation of the cost function substantially. The circuit no longer has to check whether certain combinations of qubits are in the same or opposite states, thus avoiding costly CNOT gates. It instead adds all edge costs the encoding qubit of which is in a $\vert1\rangle$ state. In our implementation this is done using phase gates where edge costs are encoded in the phase. Applying this logic to all qubits has the effect that in the tensor product those phases are summed up. Consequently, one has to make sure that the phase sum does not exceed $2\pi$ for the result to remain meaningful. We address this by normalizing all edge costs prior to building the circuit. \\

The standard TSP encoding for QAOA exhibits cyclic redundancies. A tour $(\iota_1, ... ,\iota_{n-1}, \iota_n)$ is identical to the tour cyclically shifted by one index $(\iota_n, \iota_1, ... \iota_{n-1})$. The qubit representation however is different. In noisy environments, due to those redundancies, measurements of feasible solutions are prone to be drowned out by occurrences of infeasible states induced by noise, since all tour probabilities are distributed across their cyclic variants. Such behavior can be remedied by fixing one vertex as the start point of any tour, which incidentally reduces the number of required qubits by $2n - 1$. 

\begin{figure}[htbp]
\centerline{\includegraphics[width=0.3\textwidth]{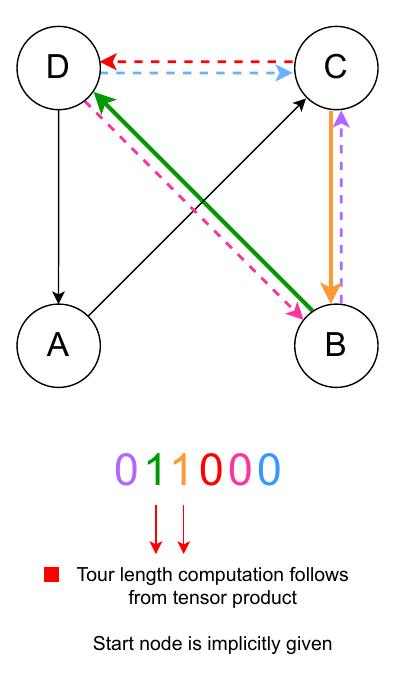}}
\caption{The same tour as in Figure \ref{EncodingOldSolo} encoded as a selection of edges. Node $A$ is given as the start node such that the inclusion of edges $e_{A,C}$ and $e_{D,A}$ can be easily derived.}
\label{EncodingNewSolo}
\end{figure}

For our encoding, we similarly and without loss of generality fixed $v_0$ as the starting and end point of any tour. Edge costs to and from $v_0$ are implicitly included by partially adding those costs by default and subtracting them when an edge is positively not part of the tour. An example is illustrated in Figure \ref{EncodingNewSolo}. \\

Let $$x_{j,k}, \ 0 \leq j,k \leq n-1, \ j \neq k$$ be binary decision variables that express whether an edge $e_{j,k}$ with cost $c(e_{j,k})$ is part of a tour, where the total tour cost amounts to 
\begin{align}
    C(x) = \sum_{j,k} x_{j,k}\cdot c(e_{j,k})
\end{align}

We reduce the number of decision variables by including edge costs from and to the start vertex $v_0$ by default and subtracting them for every conflicting edge. If an edge $e_{j,k}$ is part of a tour, neither $e_{j,0}$ nor $e_{0,k}$ can be in the tour, since the outgoing edge from $v_j$ as well as the incoming edge for $v_k$ are given by $e_{j,k}$. Hence, we can reduce the range of $i,j$ to $1 \leq j,k \leq n-1$ and update the cost function as follows:

\begin{align}
\begin{aligned}
   C(x) = & \sum\limits_{j,k} (\ c(e_{j,0}) + c(e_{0,k})\ )\ + \\ 
    & \sum\limits_{j,k} x_{j,k}\cdot (\ c(e_{j,k}) - c(e_{j,0}) - c(e_{0,k})\ )
\end{aligned}
\end{align}

For a graph with $n$ vertices $(n-1)\cdot(n-2) \in \mathcal{O}(n^2)$ such decision variables remain. \\

In the case where an edge is not part of the tour (i.e. $x_{j,k} = 0$), a fraction of the edge cost $\frac{c(e_{j,0})}{n-2}$ as well as $\frac{c(e_{0,k})}{n-2}$ is added to the total tour cost such that when $\sum_j x_{j,k} = 0$ for some $1 \leq k \leq n-1$, a total cost of $(n-2)\cdot \frac{c(e_{0,k})}{n-2} = c(e_{0,k})$ is added onto the tour, i.e. if $v_k$ is not approached by any $v_j, \ 1 \leq j \leq n-1$, it must be approached by $v_0$, since all vertices must be part of the tour. The same argument is made when $\sum_k x_{j,k} = 0$ for some $1 \leq j \leq n-1$. \\

If an edge is part of the tour (i.e. $x_{j,k} = 1$), the edge cost $c(e_{j,k})$ for that edge must be added to the total tour cost. Since $\sum_{l\neq j} x_{l,k} = 0$ and $\sum_{l\neq k} x_{j,l} = 0$ due to TSP boundary conditions, the partial edge costs $(n-3)\cdot\frac{c(e_{j,0})}{n-2}$ and $(n-3)\cdot\frac{c(e_{0,k})}{n-2}$ that are added through those decision variables need to be subtracted from the tour cost. This is because $e_{j,0}$ can not be part of the tour when $e_{j,k}$ already is the outgoing edge from $v_j$. A similar argument can be made for $e_{0,k}$ which can not be the incoming edge for $v_k$ in that case. \\

We encode both cases in a phase gate:

\begin{align}
    & P_{x_{j,k}=1} \cdot X \cdot P_{x_{j,k}=0} \cdot X \\
    & \begin{aligned}
        & = \begin{pmatrix} 
                1 & 0 \\ 
                0 & e^{i\cdot (c(x_{j,k}) - (n-3)\cdot\frac{c(e_{j,0})}{n-2} - (n-3)\cdot\frac{c(e_{0,k})}{n-2})}
            \end{pmatrix} \cdot \\ 
        & \ X \cdot
            \begin{pmatrix}
                1 & 0 \\ 
                0 & e^{i\cdot (\frac{c(e_{j,0})}{n-2} + \frac{c(e_{0,k})}{n-2})} 
            \end{pmatrix}\cdot X \\
    \end{aligned} \\
    % & = \begin{pmatrix}
    %             e^{i\cdot (\frac{c(e_{j,0})}{n-2} + \frac{c(e_{0,k})}{n-2})} & 0 \\
    %             0 & e^{i\cdot (c(x_{j,k}) - (n-3)\cdot\frac{c(e_{j,0})}{n-2} - (n-3)\cdot\frac{c(e_{0,k})}{n-2})}
    %         \end{pmatrix} \\
    & = P_{j,k}
\end{align}

Since this is a diagonal operator the operator on the entire qubit array is diagonal as well. Let $0 \leq a \leq 2^{(n-1)(n-2)}$ be an integer the binary representation of which describes a valid TSP tour. We use $a$ as an index in the diagonal operator:

\begin{align}
    & \left( \bigotimes_{j,k} P_{j,k} \right)_{a,a} \\
    & \begin{aligned}
        & = \prod_{x_{j,k}\in a,\ x_{j,k} = 0} e^{i\cdot (\frac{c(e_{j,0})}{n-2} + \frac{c(e_{0,k})}{n-2})} \cdot \\ 
        & \ \prod_{x_{j,k}\in a,\ x_{j,k} = 1} e^{i\cdot (c(x_{j,k}) - (n-3)\cdot\frac{c(e_{j,0})}{n-2} - (n-3)\cdot\frac{c(e_{0,k})}{n-2})}
    \end{aligned} \\
    & \begin{aligned}
        & = exp(\ i\cdot (\sum\limits_{x_{j,k}\in a} ( (1 - x_{j,k})\cdot (\frac{c(e_{j,0})}{n-2} +\frac{c(e_{0,k})}{n-2})\ +  \\
        & \ x_{j,k}\cdot (c(e_{j,k}) - (n-3)\cdot\frac{c(e_{j,0})}{n-2} - (n-3)\cdot\frac{c(e_{0,k})}{n-2})))) 
    \end{aligned} \\
    & \begin{aligned}
        & = exp(\ i \cdot (\sum\limits_{x_{j,k}\in a} ( \frac{c(e_{j,0})}{n-2} + \frac{c(e_{0,k})}{n-2} + \\
        & \ x_{j,k}\cdot (\ c(e_{j,k}) - c(e_{j,0}) - c(e_{0,k})\ ))))
    \end{aligned} \\
    & \begin{aligned}
        & = exp(\ i\cdot (\sum\limits_{j,k} (\ c(e_{j,0}) + c(e_{0,k})\ )\ + \\
        & \ \sum\limits_{x_{j,k}\in a} x_{j,k}\cdot (\ c(e_{j,k}) - c(e_{j,0}) - c(e_{0,k})\ )))
    \end{aligned}
\end{align}

Hence, for an individual encoded tour $a$ the correct tour cost is expressed in the operator phase, satisfying the QAOA requirements for a phase separator. \\

\section{Experiments} \label{experiments}

We compare the implementation of Grover QAOA in \cite{gqaoa} with an implementation of our proposed encoding using Grover mixers. We randomly generated 1000 TSP instances of size $n=4$. All edge lengths are integer values sampled uniformly from $[1,..,20]$, hence our random instances are in general neither symmetric nor do they obey the triangle inequality. All hereby generated weight matrices were normalized with $n\cdot max$ before they were applied to the circuits such that the phase value in the phase separator can not exceed $2\pi$. All QAOA circuits were optimized using COBYLA.

Our Qiskit implementation is available at \footnote{\label{repo}\url{https://github.com/ekut-es/qaoa-encoding}}. Note that this implementation is static in the given problem size and only offers solvers for TSP instance sizes $3$ and $4$, since the circuit examples given in \cite{gqaoa}, which we want to compare to, cover only those instance sizes as well. We chose to simulate both QAOA variants noiselessly due to their vastly different qubit and gate counts, which would drastically favor our proposed circuit in a noisy environment. This would prevent deriving meaningful conclusions in regards to the merits of each tour encoding on its own. Our code also includes an implementation of the Held-Karp algorithm to optimally solve TSP instances, allowing us to compare QAOA solutions to their respective optima. \\

\begin{figure}[htbp]
\centerline{\includegraphics[width=0.5\textwidth]{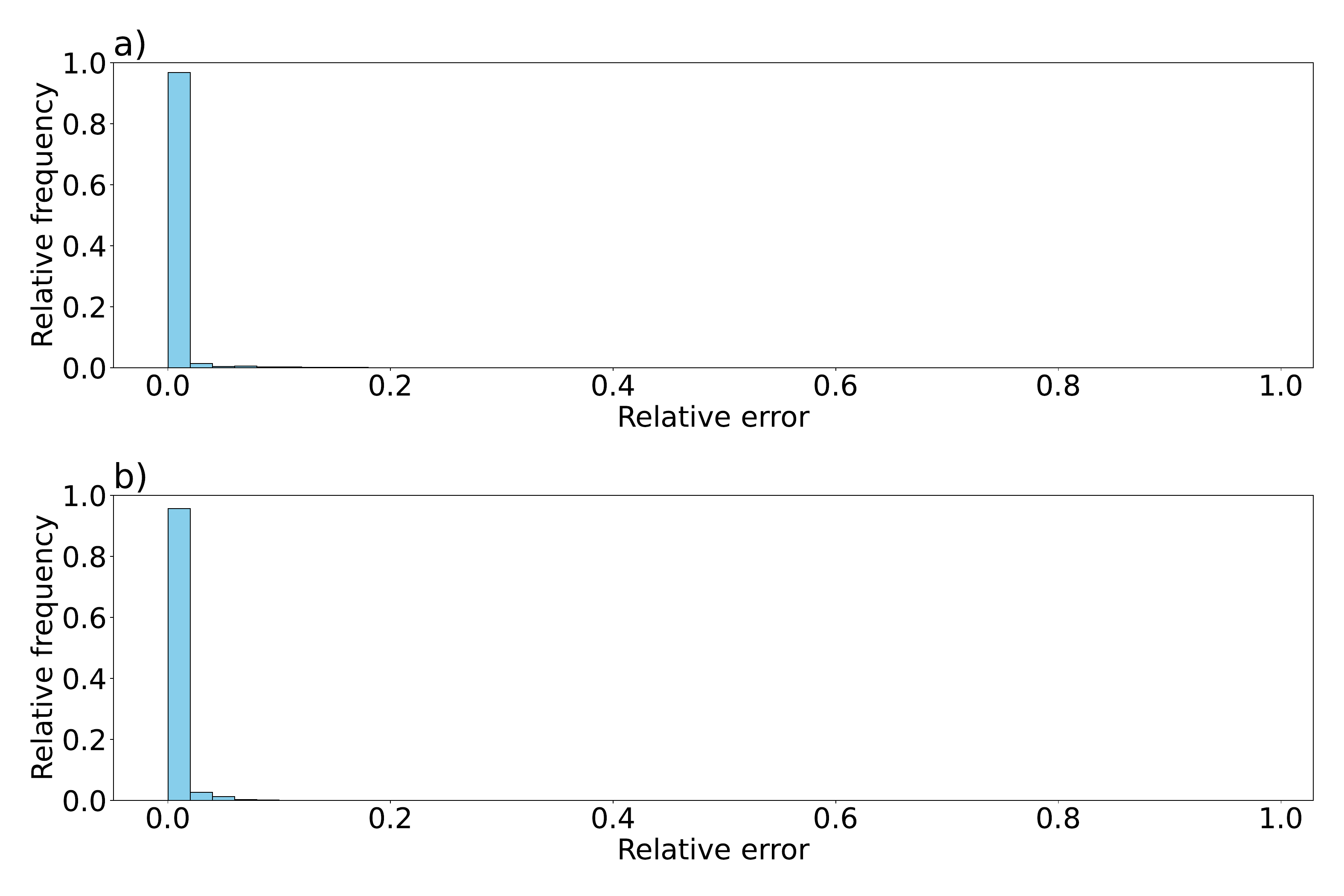}}
\caption{Relative error distributions over 1000 randomly generated TSP instances with 4 vertices for the conventional QAOA encoding (top) and our proposed encoding (bottom)}
\label{error_dist}
\end{figure}

We first examine the approximation ratio. Figure \ref{error_dist} features the relative error distribution for both the conventional (top) and the proposed encoding (bottom). Note that a relative error of $0$ corresponds to an approximation ratio of $1$. It is immediately obvious that the circuit using our proposed encoding finds the optimal solution just as well as the one using the conventional encoding. The distribution shapes are also very alike, suggesting that the difficulty of finding a solution does not change. The average relative error is $0.439$ for the conventional encoding and $0.042$ for ours. \\

\begin{figure}[htbp]
\centerline{\includegraphics[width=0.5\textwidth]{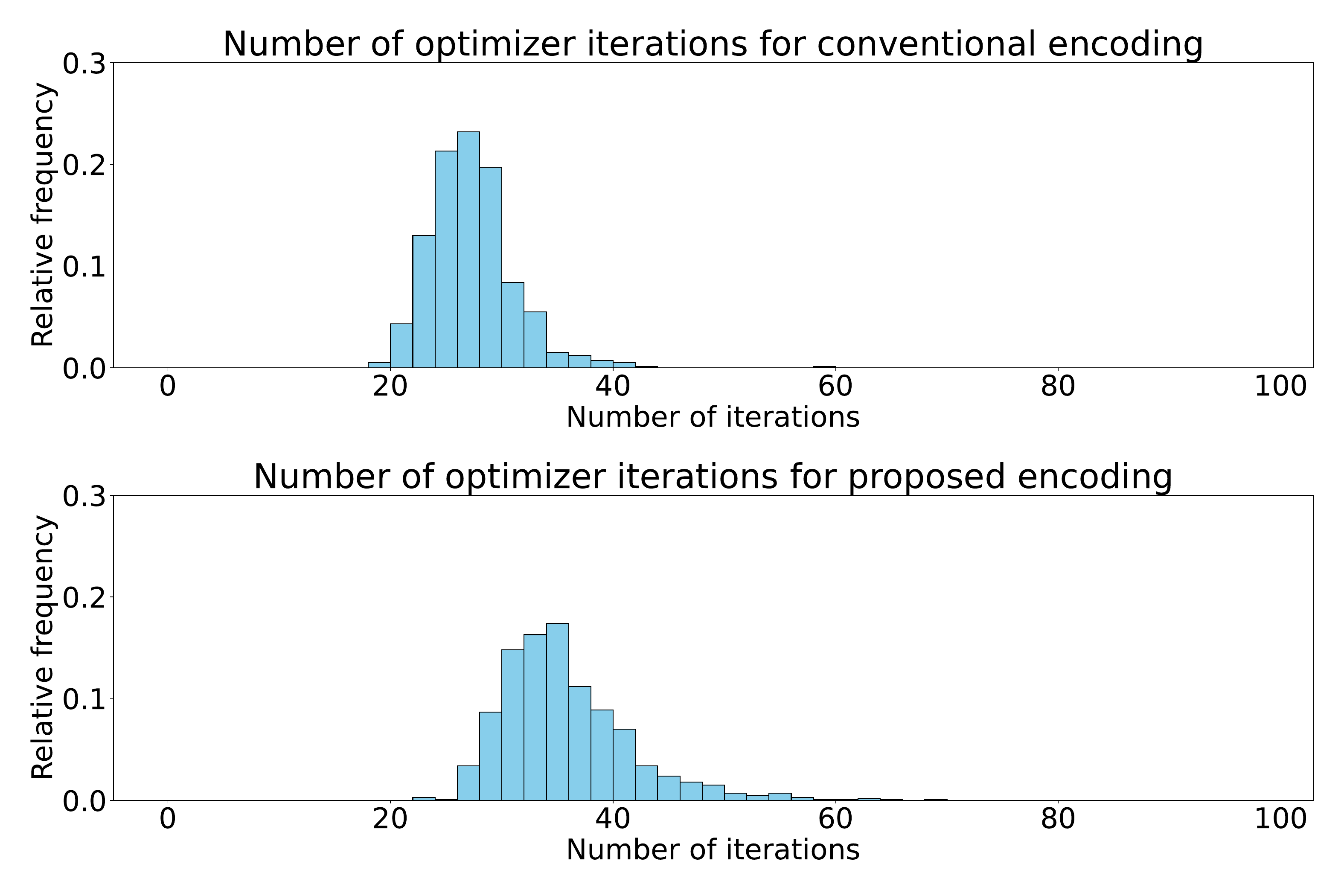}}
\caption{Distributions of the number of optimization steps required for 1000 randomly generated TSP instances with 4 vertices for the conventional QAOA encoding (top) and our proposed encoding (bottom)}
\label{optim_dist}
\end{figure}

Next we inspect the number of optimizer iterations that were necessary for QAOA to converge. In Figure \ref{optim_dist} the distributions of the number of required optimizer iterations are shown. While both distributions have a similar shape, the QAOA variant with the conventional encoding converges noticeably quicker than our proposed one. The average amount of iterations is $26.778$ for the former and $35.261$ for the latter.

\section{Discussion}

We presented an alternative problem encoding for the TSP to use in a QAOA environment and provided a mathematical proof of its correctness. Our proposed encoding requires fewer qubits than the conventional one and removes cyclic redundancies in its expressiveness. In experiments we showed that QAOA circuits using our proposed encoding return comparable results at the cost of more optimizer iterations.

The encoding proposed in this paper exhibits even more advantages in TSP statements where edge costs are symmetric. While the conventional encoding would not change in such an instance, with ours the number of required qubits would be halved, since the number of unique edges would be halved as well.

We did not provide a generic construction formula for the mixer circuit, which remains a missing link at the point of writing. If one would want to implement Grover mixers due to their computational advantages, they would have to find a circuit that generates a balanced superposition of all possible TSP routes. This might prove to be difficult, even though it can be done for the conventional encoding \cite{gqaoa}. A different approach could be similar to the one in \cite{qaoap} where only local transformations to a valid tour are made, such as flipping an edge as well as all adjacent edges. Our results show that finding such a mixer could prove advantageous in making a convincing argument for QAOA approximations as a valid alternative to classical ones.

\bibliography{bibliography}

\end{document}